# Preparation of gold-modified palladium thin films as efficient catalysts for oxygen reduction reaction

Fereshteh Dehghani Sanij [a], Vitalii Latyshev [a], Serhii Vorobiov [a], Hoydoo You [b], Dominik Volavka [a], Tomas Samuely [a], and Vladimir Komanicky [a*]

[a] Institute of Physics, Faculty of Science, P. J. Safarik University, Park Angelinum 9, Kosice 040 01, Slovak Republic.

[b] Materials Science Division, Argonne National Laboratory, Argonne, Illinois 60439, USA.

*Corresponding author:*

E-mail address: vladimir.komanicky@upjs.sk (V. Komanicky).

**Abstract**

Devising effective platinum (Pt)-free nanomaterials for catalyzing oxygen reduction reaction (ORR) is desired, but some main drawbacks still exist for commercial implementation in fuel cell systems. Herein, we present the preparation of gold (Au)-modified palladium (Pd) thin films with various thicknesses of Au using magnetron sputtering deposition. It is found that the prepared Au-modified Pd (Au/Pd) thin films with some desirable characteristics such as large electrochemical surface area (ECSA), improved electron transfer, along with high catalytical active positions reveal markedly boosted electro-catalytic activity and stability for oxygen reduction in acidic environments compared with pure Pd catalyst. Noticeably, the as-deposited Au/Pd thin films exhibit a connection in oxygen reduction ability as a function of Au thickness, with the 0.16 nm Au/Pd thin film being the best ORR sample. Significantly, 0.16 nm Au/Pd material demonstrates superior stability after 1000 potential cycles in comparison with the 0.16 nm Pt/Pd owing to the influence of Au modification as well as its structurally stable surfaces.

## 1. Introduction

Oxygen reduction reaction (ORR) is one of the most widely investigated electro-catalytical procedures since this half-reaction is a major issue restricting the efficiency of fuel cell technologies [1,2]. Platinum (Pt)-based nanostructured catalysts have been generally utilized as cathodic materials in such devices, whereas their rarity and high price prohibit the broad-based industrial applications of fuel cell systems [3,4]. Hence, designing advanced non-Pt electro-catalysts for ORR is still a crucial hurdle in fuel cells.

Palladium (Pd), amongst the various metallic elements under scrutiny, possesses characteristics closely analogous to those of Pt, although being more available and significantly less expensive [5–7] as the number of internal combustion engines decreases. Recently, the oxygen reduction on Pd and its alloyed nanocatalysts has attracted considerable interest [8,9]. Pd-based materials are the electro-catalysts with the second-best catalytic activity for ORR, behind Pt catalysts. The catalytic selectivity towards the ORR process in the existence of methanol in direct methanol fuel cells (DMFCs) is another benefit of Pd-based catalysts [10,11]. Numerous studies have suggested that Pd-based alloys and core-shell nanostructures may have superior oxygen reduction catalytic properties compared with Pd alone, which could be associated with the adjustments of d-band position and lattice distortion [9,12]. Nevertheless, under operating environments for the cathodic nanocatalysts of fuel cells, the 3d transitional metal components may dissolve into acidic solutions.

The electrochemical stability of the catalysts is just as essential for the industrialization of fuel cell systems as their catalytic activity and price. As a consequence of the surface metallic atoms dissolving and accumulating in complex electrochemical circumstances, the normally inert metal becomes unstable [13,14]. In addition, earlier investigations have shown that introducing other metallic components into nanostructures can modify electronic constructions and increase stability over oxygen reduction procedures [15–17]. Gold (Au) has so far been demonstrated as a durable element that enhances the stability of Pt catalysts towards oxygen reduction as a result of its much lesser oxophilicity relative to Pt [18]. For instance, the deposition of Au clusters on the surface of nanoparticulated Pt catalysts has been described to stabilize the Pt nanocatalysts against dissolution during accelerated stability tests owing to diminishing PtOH generation and consequently boosting their oxygen reduction durability [19]. The stability enhancements of Pt materials for ORR through partly covering the surface of catalysts by Au atoms utilizing a variety of techniques have been stated in previous works by several research teams [20–22]. Up to now, very few scientists have concentrated on the synergetic interaction between Au and metallic nanostructures for concurrently increasing the electro-catalytic activity and durability of Pd-containing materials for catalyzing oxygen reduction.

In the present work, we developed Au-modified Pd (Au/Pd) thin film catalysts with different thicknesses of Au through the magnetron sputtering deposition method, intending to improve durability without reducing electro-catalytic activity. The activities of these thin films for ORR catalysis were examined by some electrochemical methods. Although bulk Au seems to be inactive, the 0.16 nm coverage of Au (0.16 nm Au-modified Pd thin film) specified noticeably higher electro-catalytic ability towards ORR compared to other studied samples. As far as we know, no previous study has been published showing the influence of such Au coverage on Pd electrode activities for oxygen reduction in acidic media. The optimum Au-modified Pd (0.16 nm Au/Pd) thin film exhibited similar behavior to the Pt catalyst towards oxygen reduction but more intriguingly possessed much greater durability compared with 0.16 nm Pt-modified Pd (Pt/Pd) thin film in acidic conditions. Moreover, physical analyses of the thin films indicate that the thickness of the Au can substantially affect the ORR electro-catalytic properties. The augmented activity and stability of the optimum Au/Pd samples can be attributed to Au modification, the synergistic effects, along with unique electronic effects observed in binary thin film catalysts.

## 2. Experimental section

### 2.1. Sample fabrication

Pt, Pd, Au, and Au-modified Pd thin films were deposited on the surface of glassy carbon (GC) electrodes using the magnetron sputtering method (Orion 8, AJA International sputtering system) from Pt (99,98% purity), Pd (99,98% purity), and Au (99,98% purity) targets at room temperature (Fig. 1). The sputtering deposition was conducted at a pressure of 3 mTorr in the Ar atmosphere with a purity of 99.999%, while the base pressure within the vacuum chamber was around $10^{-7}$ Torr. The thicknesses of the deposited Pt, Pd, and Au films were 20 nm. The Au-modified Pd thin films were deposited in a sequential manner-Au on top of 20 nm Pd film. The deposition rate of Au was 0.8 Å s$^{-1}$ at 25 W of the discharge power. To obtain the various thicknesses of Au in Au/Pd thin films, the deposition time of Au was controlled. To assure the repeatability of the deposition, the parameters of the preparation procedure were controlled by computer.

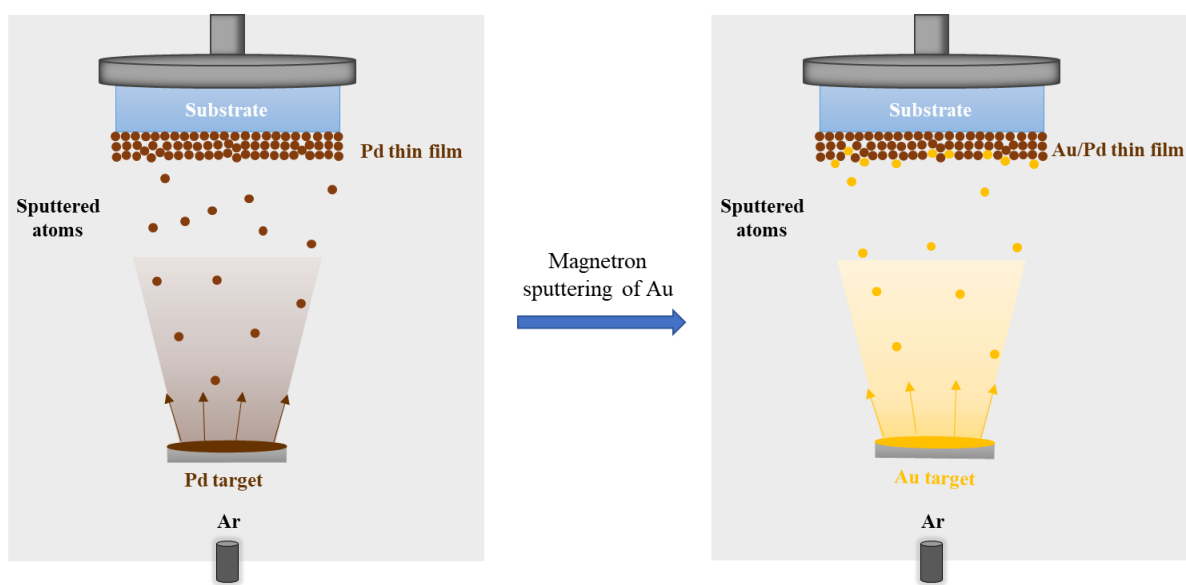

**Fig. 1.** Schematic illustration of magnetron sputter deposition of Au-modified Pd thin film catalysts.

*2.2. Physical characterization*

The surface morphology of the thin films was analyzed through atomic force microscopy (AFM, Bruker, ICON). The AFM images were recorded in tapping mode using silicon tips under ambient conditions. The X-ray photoelectron spectroscopy (XPS) measurements were performed by the high-resolution spectrometer SPECS PHOIBOS 100/150 using an Al anode with a power of 200 W. The pressure during measurement was $10^{-8}$ mbar. The XPS experiments were conducted at room temperature. The investigated samples were placed on a molybdenum sample holder using conductive carbon tape. All Au-modified Pd and unmodified Pd samples were also characterized by X-ray reflectivity scans. The X-ray measurements were performed with a Rigaku 12 kW rotating anode X-ray generator equipped with a Huber four-circle diffractometer. A bent graphite (002) monochromator was used to select the Cu Kα fluorescence line and focus it onto the sample position at the center of the diffractometer with a focal distance of 50 cm.

*2.3. Electrochemical characterization*

The electro-catalytic activities of the thin film nanocatalysts were examined using a potentiostat (Autolab PGSTAT 302 N) coupled with a standard three-electrode electrochemical configuration. The Pd-based thin film catalysts were investigated as cathodes for ORR in acidic (0.1 M $HClO_4$) circumstances at room temperature. A glassy carbon electrode (GC, diameter: 6 mm) was used as the working electrode, and Pt wire and saturated Ag/AgCl (3 M NaCl) were used as the counter and reference electrodes, respectively. To prevent the acid solution from chloride contamination, the reference electrode was situated in a separate compartment linked to the primary electrochemical cell through a salt bridge. In all experiments, ultrapure (18 MΩcm) deionized (DI) water (Elga Purelab Ultra) was employed to make aqueous solutions. The cyclic voltammetry (CV) measurements were accomplished in solutions saturated with Ar at the potential range of -0.2 V~1 V and the scan speed of 50 mV/s. The ORR experiments were conducted by utilizing linear sweep voltammetry (LSV) in $O_2$-purged 0.1 M $HClO_4$ solution at 10 mV/s with the electrode rotating at 1600 rpm. The accelerated durability tests were performed between 0.3 and 0.7 vs. Ag/AgCl (0.6 and 1.0 V vs. RHE) for 1000 cycles at 50 mV/s in Ar-purged 0.1 M $HClO_4$ solution.

## 3. Result and discussion

### 3.1. Physical characterization of the samples

The AFM images of the pure Pd and 0.16 nm Au/Pd surfaces are illustrated in Fig. 2. The AFM images show the smooth surface of the obtained thin films (Figs. 2a and b). As depicted in Fig. 2a, the average roughness ($r_q$) of the Pd surface is found to be 0.181 nm. The average roughness of the Au/Pd surface is estimated to be 0.244 nm (Fig. 2b), signifying a notable increase when compared to the pure Pd thin film.

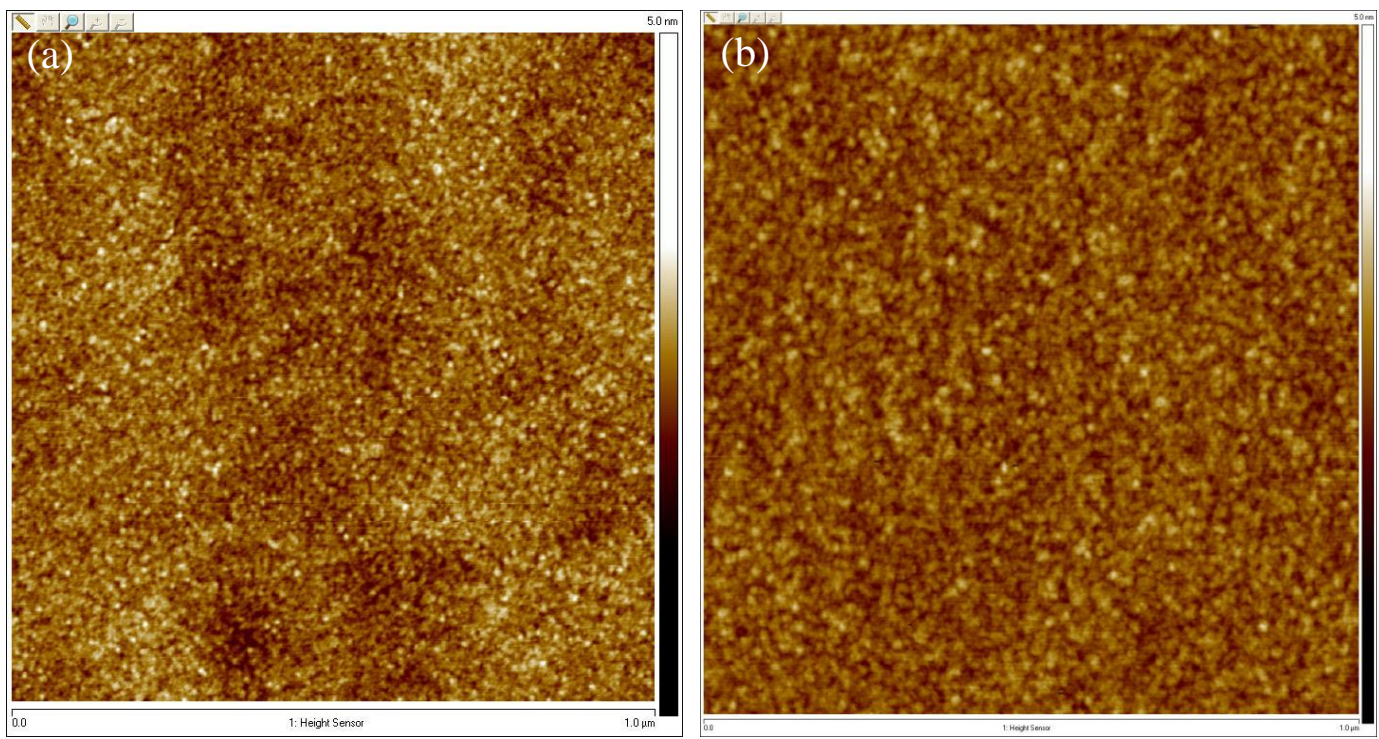

**Fig. 2.** AFM images of the pure Pd (a) and 0.16 nm Au/Pd (b) thin film samples.

XPS was employed to elucidate the surface composition of the obtained Pd, Au, and Au-modified Pd materials (Figs. 3a and b). Fig. 3a displays local Au 4f spectra of pure Au and Au-modified Pd catalysts. The pure Au spectrum exhibits a pair of spin-orbit doublets at a low-energy band and a high-energy band at 84 and 87.6 eV indexed as Au $4f_{7/2}$ and Au $4f_{5/2}$, respectively [23]. These peak locations match the outcomes described previously, signifying the presence of metallic $Au^0$ [24]. The Au 4f spectra of Au/Pd samples are also shown in Fig. 3a. It can be seen that $Au^0$ is the main species in the investigated Au/Pd materials. As presented in Fig. 3a, the Au 4f binding energies of the $Au^0$ species shifted negatively in the Au/Pd materials comparing to the locations revealed by the $Au^0$ species in pure Au. Remarkably, the negative shift of binding energies of the Au 4f peaks of Au/Pd samples increased with increasing the thickness of Au for the as-prepared 0.5 nm Au/Pd catalyst. The decrease of the binding energies of Au for the Au/Pd materials compared to the pure Au indicates the charge transfer from Pd to Au, which can be associated with the electronic interplay between atomic orbitals of metallic elements and, therefore, the alloyed structure fabrication [23,25].

Fig. 3b shows the XPS spectra of the Pd in pure Pd and Au-modified Pd catalysts. The pure Pd spectrum exhibits two peaks at a low-energy band and a high-energy band at 335.2 and 340.5 eV that belong to Pd $3d_{5/2}$ and Pd $3d_{3/2}$, respectively, demonstrating that Pd is present predominantly in metallic $Pd^0$ form [24]. The results further reveal that $Pd^0$ is dominant in the Au/Pd catalysts studied. In comparison to pure Pd, there is no obvious shift of Pd 3d binding energies for the Au/Pd catalysts; nevertheless, the spectra's shapes are slightly different, representing the changes in the composition of Au-modified samples. The Au-induced change of the electronic structure of Pd can give rise to variations in the d-band position. In addition, the surface activity of modified Pd samples is affected by changes in d-band positions, too.

Interestingly, the XPS spectra of Au and Pd for the 0.16 nm Au/Pd sample remained unchanged after 5 CV cycles, demonstrating no alternation in the catalyst's surface (Figs. 3a and b). Accordingly, 0.16 nm Au/Pd possesses improved catalytic capability relative to that of pure Pd catalysts.

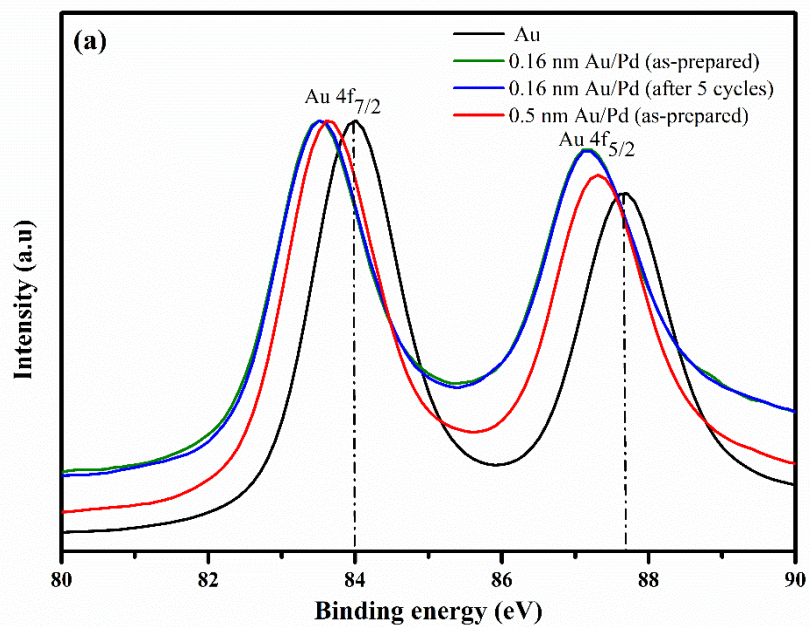

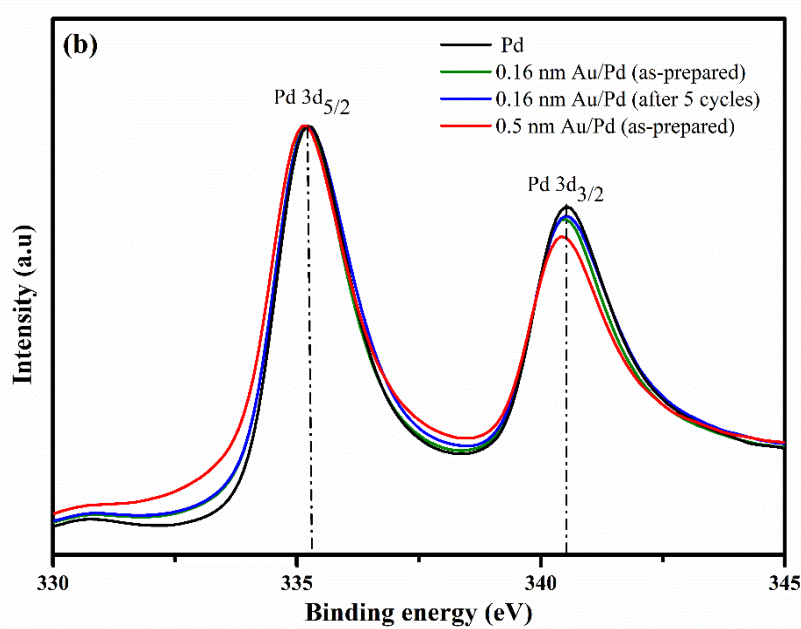

**Fig. 3.** XPS spectra of pure Pd and Au/Pd samples in the Au 4f (a) and Pd 3d (b) regions.

The thicknesses measured with X-ray reflectivity scans were found to be consistent with the thickness of Pd and Au layers expected from the calibration of sputter deposition. In Fig. 4b, the two reflectivity scans of the unmodified 20 nm thick Pd film (blue) and the sample modified with 0.16 nm thick Au (red) are shown. The blue and red solid lines fit the reflectivity scans, respectively. The normalized reflectivity is also shown in Fig. 4a, demonstrating that Au 0.16 nm is thick enough to significantly modify the reflectivity. In Fig. 4c, the respective density profiles are shown, where the thickness of Au is clearly identified. However, the electron density of the Au film from the fit is significantly lower than that of the ideal Au density, the thickness of 0.20(1) nm from the fit is larger than 0.16 nm expected from the sputtering, and the roughness of the surface increases from 0.18(1) nm to 0.28(1) nm. These X-ray reflectivity results indicate that Au does not uniformly cover the Pd surface, resulting in a reduced Pd surface exposed to the electrolyte. These results are also consistent with the AFM results discussed above.

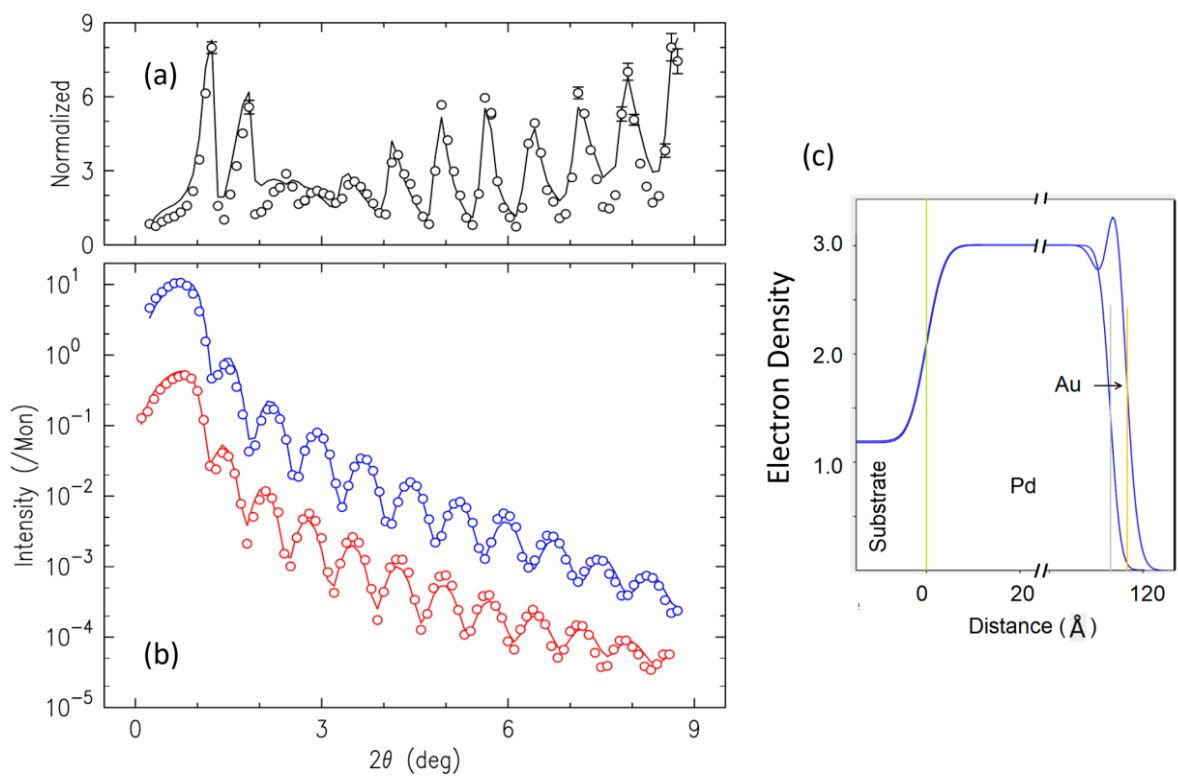

**Fig. 4.** X-ray reflectivity scans for the samples with Pd only and Pd with the thinnest layer of 0.16 nm Au. The divided reflectivity to emphasize the difference between the two reflectivity scans (a), the reflectivity scans for the Pd only and Pd with 0.16 nm Au deposited with an offset for clarity (b), and the electron density profile deduced from the reflectivity fits (c). The roughness of the Pd, the roughness of the Au/Pd, and the Au thickness obtained from the fits are 0.18(1) nm, 0.28(1) nm, and 0.20(1) nm, respectively.

*3.2. Electrochemical characterization of the samples*

*3.2.1. Cyclic voltammetry (CV) and rotating disk electrode (RDE)*

The electrochemical behaviors of prepared catalysts were assessed using cyclic voltammetry (CV) in Ar-saturated 0.1 M $HClO_4$ electrolytes. To compare the electro-catalytic ability, the CV of the pure Pt catalyst is also presented in Fig. 5. As can be seen in Fig. 5, the CV features for the modified Pd samples are similar to pure Pd, illustrating the regions of desorption/adsorption of hydrogen and formation/reduction of Pd oxide. The deposition of the higher thickness of Au considerably reduced the charge of surface oxidation/reduction for the Au-modified samples. Besides, further analysis of the CV profiles showed that Au-modified Pd thin films exhibited less anodic characteristics compared to Pd within the high potential areas associated with surface oxidation. This observation clarifies that modifying Pd catalysts with Au can be directed to better tolerance to surface oxidation in oxygen reduction catalysis.

The electrochemical active surface area (ECSA) of Pd-based catalysts was estimated by measuring the Coulombic charge for PdO reduction, divided by the theoretical charge value for the reduction of PdO monolayer (424 µC cm$^{-2}$) [26]. The ECSA of the Pt thin film sample was experimentally determined by integrating the charge corresponding to the hydrogen adsorption/desorption observed in the CV profile (Fig. 5), utilizing a conversion factor of 210 µC cm$^{-2}$ [27]. The values of calculated ECSAs are tabulated in Table 1. As expected, the ECSA value of Au-modified Pd materials decreased with the increase in the thickness of Au. The larger ECSA of the 0.16 nm Au/Pd catalyst compared to other samples with thicker layers of Au would offer a greater number of active points for the electro-catalysis of oxygen reduction.

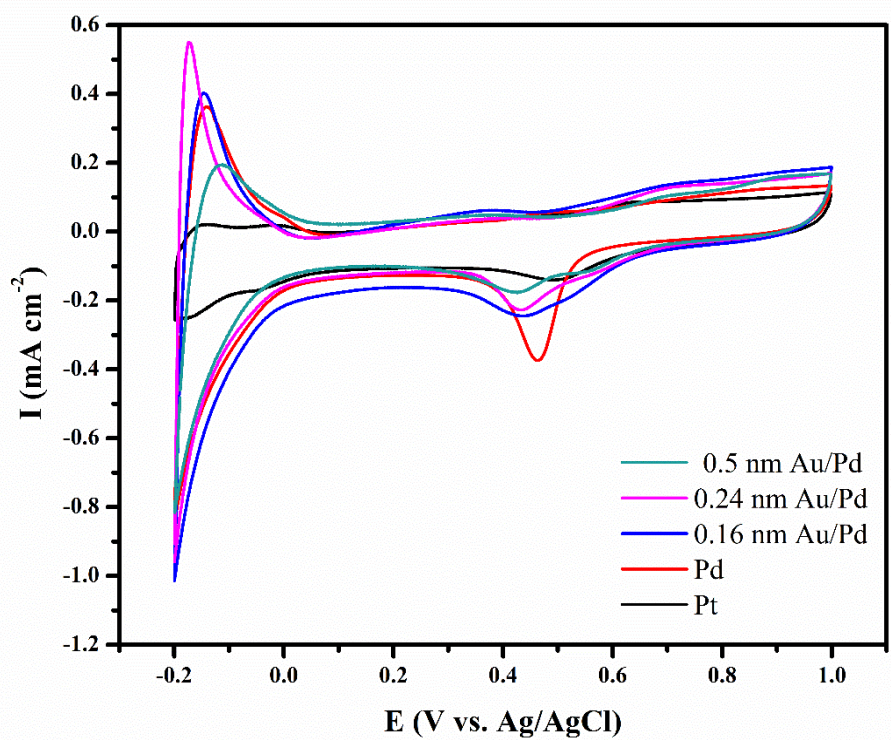

**Fig. 5.** CVs of the prepared thin film catalysts in Ar-saturated 0.1 M HClO$_4$, Potential scan rate: 50 mV s$^{-1}$.

The ORR electro-catalytic activities of all samples were evaluated in $O_2$-purged 0.1 M $HClO_4$ solutions. Fig. 6a shows the linear sweep voltammetry (LSV) curves at 1600 rpm from the rotating disk electrode (RDE) for prepared thin films. The onset potentials ($E_{onset}$) measured in 0.1 M $HClO_4$ for the resultant thin film materials are provided in Table 1. The oxygen reduction polarization curve of 0.16 nm Au/Pd thin film material unveiled higher $E_{onset}$ (0.58 V), which was close to that of pure Pt catalyst (0.60 V). It is worth noticing that on the Pt and Au-modified Pd catalysts, the mixed kinetic-diffusion control region and diffusion-controlled region in the oxygen reduction polarization curves can be clearly observed throughout the potential ranges from 0.7 to 0.3 V and below 0.3 V, respectively. Pure Au exhibited no electro-catalytic activity over the 0.7-0.3 V potential regime. Nevertheless, the 0.16 nm Au/Pd catalyst displayed much greater ORR activities compared to the pure Au sample. Moreover, it is found that the oxygen reduction ability of 0.16 nm Au/Pd catalyst is comparable to Pt despite the fact that Au atoms covered a significant portion of the Pd surface. Evidently, the ORR performance is substantially influenced by the coverage of Au on the surface of Pd catalysts.

Within the mixed potential domains, the oxygen reduction activities of various electro-catalysts can be compared using the evaluation of the half-wave potential ($E_{1/2}$). The $E_{1/2}$ values achieved from ORR polarization plots in Fig. 6a for investigated samples are compiled in Table 1. The results obviously demonstrate that the $E_{1/2}$ is shifted more positively (~28 mV) in the case of 0.16 nm Au/Pd in comparison to that acquired with the pure Pd catalyst. The prepared samples with an Au thickness in excess of 0.16 nm exhibit negative shifts in $E_{1/2}$ with a corresponding decline in the current values relative to pure Pd and Pt materials. These findings imply that as the Au thickness is increased, the ORR electro-catalytic activities of Au-modified catalysts diminish. It should be pointed out that the 0.16 nm Au/Pd catalyst behaves similarly to the pure Pt material. The obtained outcomes indicate that the sample with an Au thickness of 0.16 nm possesses an excellent ORR electro-catalytic capability. It can be supposed that an optimal thickness of Au will lead to a decrease in the generation of oxygenated species, which are identified as poisoning intermediates on the surface of Pd catalysts during the oxygen reduction process.

In order to complete the discussion, it is essential to emphasize that the Au surfaces rarely act as active oxygen reduction catalysts in acidic circumstances [28]. This is due to the fact that Au is unable to offer adsorption positions for the formation of OH $_{(ads)}$ intermediates. These intermediates, which are formed by $H_2O$ dissociating at the surface of Pd in acid solutions, could poison the electro-catalyst for ORR because their existence decreases the surface-active positions for oxygen activation through oxygen bond breaking. As previously stated, the presence of Au prevents the generation of oxygenated intermediates and thereby enables Pd to function effectively in ORR catalysis procedures.

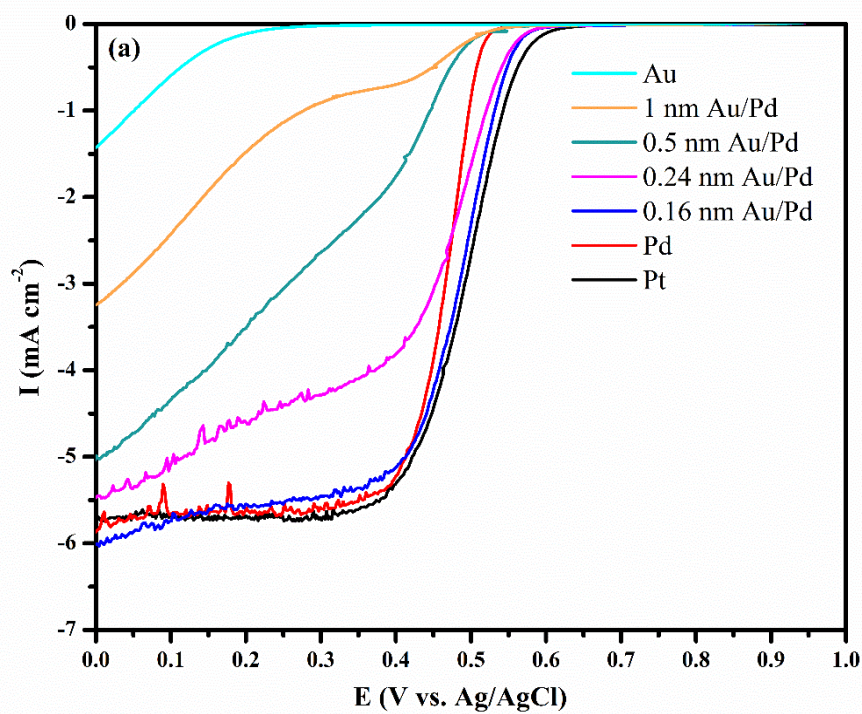

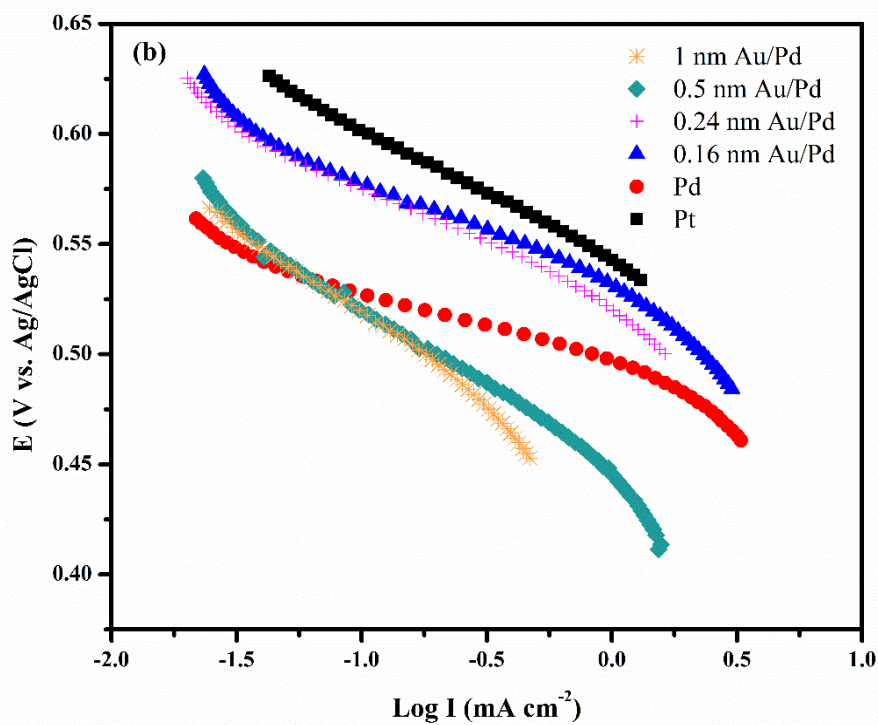

**Fig. 6.** The ORR polarization curves (a) and Tafel slope (b) of the prepared thin film catalysts in $O_2$-saturated 0.1 M $HClO_4$ at 1600 rpm, Potential scan rate: 10 mV s$^{-1}$.

To further examine the oxygen reduction kinetics catalyzed by investigated samples, the Tafel curves were plotted, as depicted in Fig. 6b. The Tafel slopes of prepared thin films are listed in Table 1. Clearly, the resulting 0.16 nm Au/Pd showed the smallest Tafel slope of 57 mV/dec, proving the highly accelerated kinetics of this catalyst in the ORR procedure.

Table 1. Electrochemical properties of the prepared thin film catalysts in 0.1 M $HClO_4$.

| Catalysts | $E_{onset}$ (V vs. Ag/AgCl) | $E_{1/2}$ (V vs. Ag/AgCl) | ECSA ($cm^2$) | Tafel slope (mV/dec) |
|---|---|---|---|---|
| Pt | 0.60 | 0.50 | 0.18 | 59 |
| Pd | 0.53 | 0.46 | 0.38 | 55 |
| 0.16 nm Au/Pd | 0.58 | 0.49 | 0.35 | 57 |
| 0.24 nm Au/Pd | 0.56 | 0.45 | 0.29 | 58 |
| 0.5 nm Au/Pd | 0.52 | 0.44 | 0.24 | 79 |
| 1 nm Au/Pd | 0.50 | 0.43 | - | 83 |

The methanol resistance behavior of Pd-containing nanocatalysts towards oxygen reduction is essential as MeOH could transfer from the anodic to the cathodic side via the membranes, poisoning the cathodic catalysts along with lessening the open circuit voltage (OCV) [11]. The LSV curves of the 0.16 nm Au/Pd and pure Pt catalysts were evaluated in $O_2$-purged solutions containing 0.1 M MeOH in 0.1 M $HClO_4$ (Figs. 7a and b). In the presence of 0.1 M MeOH, the 0.16 nm Au/Pd sample does not display any peak of methanol oxidation, demonstrating its outstanding methanol resistance (Fig. 7a). On the other hand, under identical circumstances, the pure Pt thin film catalyst illustrates an observable peak of methanol oxidation at about 0.55 V (Fig. 7b). Accordingly, it is recommended that 0.16 nm Au-modified Pd catalysts are an appropriate option for the systematic design of methanol-resistant cathodic materials for usage in DMFCs.

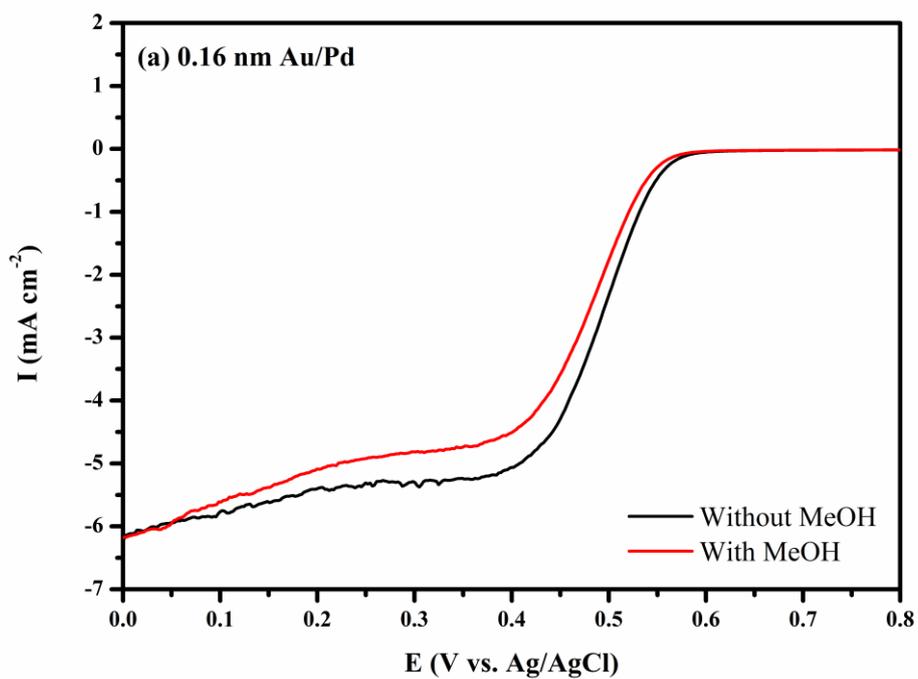

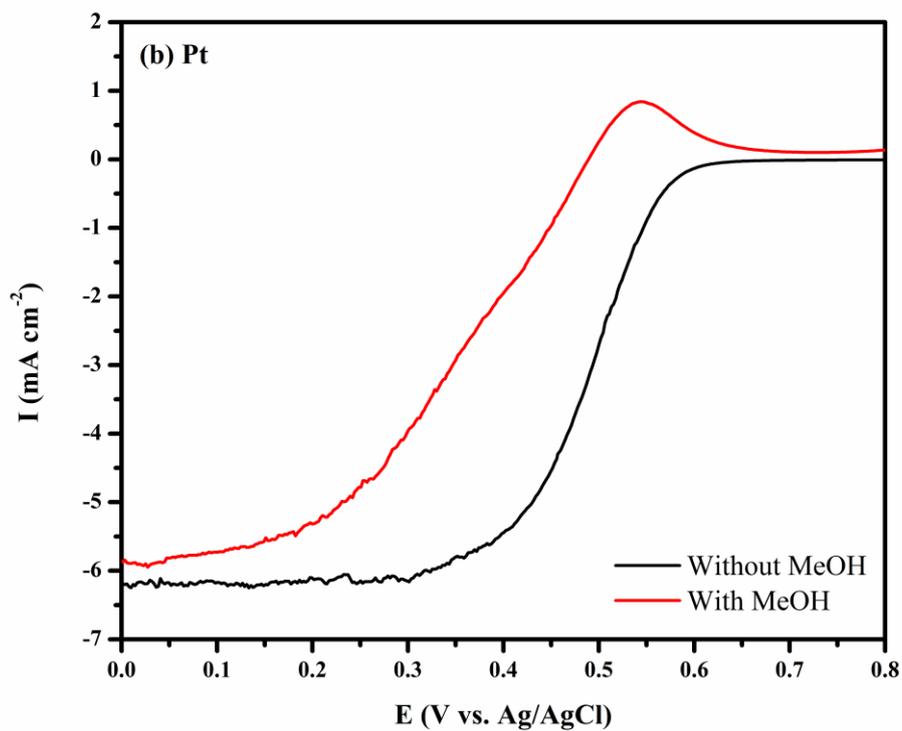

**Fig. 7.** The ORR polarization curves of the 0.16 nm Au/Pd (a) and Pt (b) thin film catalysts with and without MeOH in $O_2$-saturated 0.1 M $HClO_4$ at 1600 rpm, Potential scan rate: 10 mV $s^{-1}$.

*3.2.2. Accelerated durability test (ADT)*

Durability is a vital concern that needs to be addressed before the widespread adoption of fuel cell technologies. Thereby, an accelerated durability test (ADT) for 1000 continuous cycles at a scan rate of 50 mV/s in an $N_2$-purged 0.1 M $HClO_4$ electrolyte was done to examine the stability of these thin film materials. As revealed in Figs. 8a and b, the 0.16 nm Au/Pd illustrated superior long-term durability relative to the 0.16 nm Pt/Pd over ADTs. After 1000 continuous CV cycles, the ORR polarization curve for the 0.16 nm Au/Pd manifests a negative shift of just 10 mV in $E_{1/2}$ (Fig. 8a). In contrast, the ORR polarization curve for the 0.16 nm Pt/Pd exhibits a noticeable negative shift of around 30 mV in $E_{1/2}$, probably attributable to the gradual leaching out of the surface Pt sample after 1000 CV scans during the potential cycling (Fig. 8b). Notably, the considerable degradation after ADT measurements can lead to a drastic diminution in the oxygen reduction characteristics. This increased durability of 0.16 nm Au/Pd can be caused by the incorporation of Au and alloyed construction of the investigated sample. The improved durability of the 0.16 nm Au/Pd sample aligns with the Au influence on Pt catalysts earlier published [19,22].

As depicted in Table 2, the ORR performances of the prepared samples are compared with other catalysts documented in previous studies. The superior oxygen reduction performance of the 0.16 nm Au/Pd thin film catalyst can primarily be associated with the following aspects: the addition of Au into the Pd structure efficiently adjusts the electronic configuration of Pd due to ligand (electronic) and strain (geometric) effects, resulting in a greater number of active positions free from oxygen-containing species for catalyzing oxygen reduction [17,29]. Furthermore, the Au/Pd structure would considerably remove intermediated species generated through oxidation and markedly improve tolerance to poisoning, thereby enhancing durability [30].

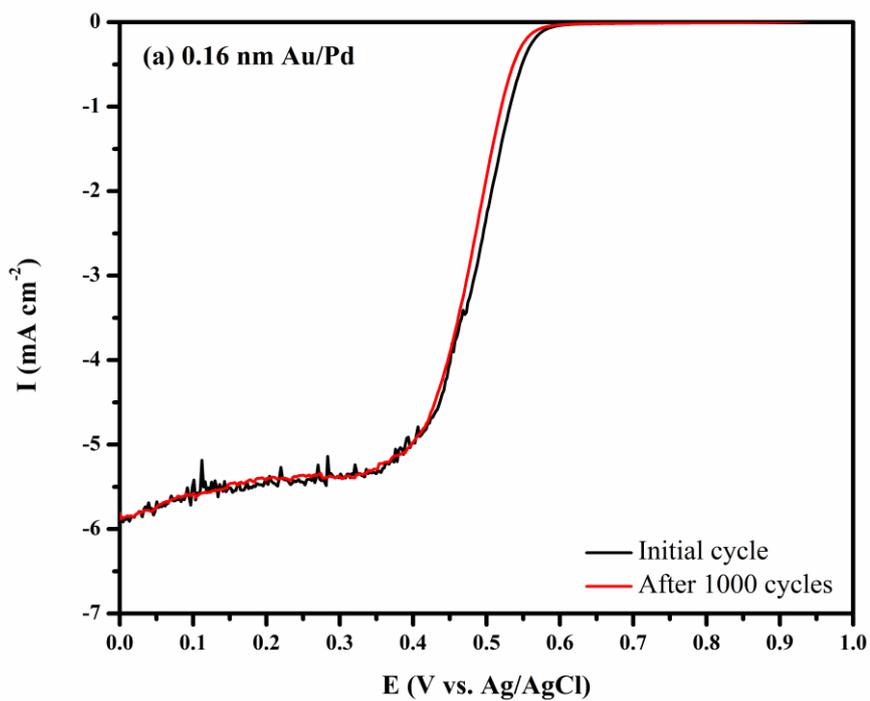

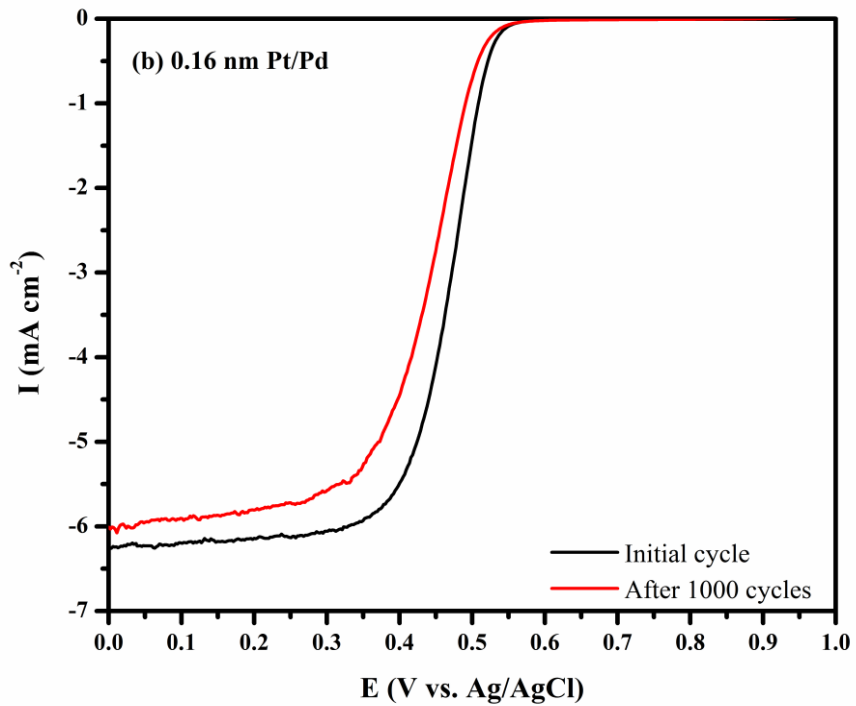

**Fig. 8.** The ORR polarization curves of the 0.16 nm Au/Pd (a) and 0.16 nm Pt/Pd (b) thin film catalysts before and after durability test in $O_2$-saturated 0.1 M $HClO_4$ at 1600 rpm, Potential scan rate: 10 mV s$^{-1}$.

In the meantime, incorporating Au can significantly modify the electronic arrangement of Pd caused by ligand and strain effects, thus giving rise to the amendment of the Pd d-band position [24,31]. This suggestion is further supported by XPS observations, which confirm the electron-withdrawing from Au to Pd resulting from the downshifting of the d-band position of Pd (Fig. 3). The optimum Au/Pd thin film can provide more active points for oxygen reduction while concurrently decreasing the ability for physisorption of oxygen and the chemisorption of oxygen-containing species produced over the electro-catalyst's surface [32]. This is in accordance with earlier investigations, underlining the synergetic interactions of two metallic constituents to boost the electro-catalytic stability of Pd [33,34].

**Table 2.** ORR catalytic performance of the prepared thin films compared with those of the previous catalysts reported in the literature.

| Catalysts | $E_{onset}$ | $E_{1/2}$ | ECSA ($cm^2$) | Tafel slope (mV/dec) | Electrolyte | Ref. |
|---|---|---|---|---|---|---|
| **Pt** | 0.60 (V vs. Ag/AgCl) | 0.50 (V vs. Ag/AgCl) | 0.18 | 59 | 0.1 M $HClO_4$ | This work |
| **Pd** | 0.53 (V vs. Ag/AgCl) | 0.46 (V vs. Ag/AgCl) | 0.38 | 55 | 0.1 M $HClO_4$ | This work |
| **0.16 nm Au/Pd** | 0.58 (V vs. Ag/AgCl) | 0.49 (V vs. Ag/AgCl) | 0.35 | 57 | 0.1 M $HClO_4$ | This work |
| **Bulk Pt** | - | 0.51 (V vs. SCE) | - | 63 | 0.1 M $HClO_4$ | [35] |
| **Bulk Pt** | - | 0.43 (V vs. SCE) | - | 69 | 0.05 M $H_2SO_4$ | [35] |
| **20 nm Pt** | - | 0.79 (V vs. RHE) | 0.46 | 86 | 0.5 M $H_2SO_4$ | [36] |
| **Poly-Pt** | 0.98 (V vs. RHE) | 0.90 (V vs. RHE) | - | - | 0.1 M $HClO_4$ | [37] |
| **5.4 nm {100}-oriented Pt/C NPs** | - | 0.81 (V vs. RHE) | 0.16 | 69 | 0.5 M H2SO4 | [38] |
| **Pt NPs** | 0.67 (V vs. Ag/AgCl) | 0.58 (V vs. Ag/AgCl) | - | - | 0.5 M H2SO4 | [39] |
| **Bulk Pd** | - | 0.77 (V vs. RHE) | - | 61 | 0.1 M $HClO_4$ | [40] |
| **10 nm Pd/Au** | - | 0.75 (V vs. RHE) | - | 61 | 0.1 M $HClO_4$ | [40] |
| **0.5 nm Pd/Au** | - | 0.59 (V vs. RHE) | - | 54 | 0.05 M $H_2SO_4$ | [40] |
| **Bulk Pd** | - | 0.71 (V vs. RHE) | 0.52 | 53 | 0.05 M $H_2SO_4$ | [41] |
| **900 s deposited Pd/Au** | - | 0.71 (V vs. RHE) | 0.40 | 47 | 0.05 M $H_2SO_4$ | [41] |
| **Au@Pd-0.2** | - | 0.52 (V vs. Ag/AgCl) | - | - | 0.1 M $HClO_4$ | [42] |
| **PdAu-500 °C-1h** | 0.98 (V vs. RHE) | 0.83 (V vs. RHE) | - | 59 | 0.5 M $H_2SO_4$ | [43] |
| **Au@Pd-I NPs** | - | 0.51 (V vs. Ag/AgCl) | - | - | 0.1 M $HClO_4$ | [44] |
| **$Au_{60}Pd_{40}$@Pt** | - | 0.94 (V vs. RHE) | - | - | 0.1 M $HClO_4$ | [45] |

## 4. Conclusions

In conclusion, a simple magnetron sputtering technique was implemented to fabricate Au-modified Pd thin film catalysts with different thicknesses of Au and investigate their catalytic performance for ORR catalysis. It has been observed that the Au-modified Pd thin film with 0.16 nm Au thickness is a remarkably active material, whilst bulk Au serves as an ineffective and oxidation-tolerant catalyst for ORR. Among the fabricated catalysts, the 0.16 nm Au/Pd sample revealed favorable performances for ORR in acidic environments with regard to outstanding electro-catalytic activity ($E_{onset}$ = 0.58 V and $E_{1/2}$ = 0.49 V) along with high durability ($\Delta E_{1/2}$, ~10 mV), and comparable to the pure Pt catalyst. The Au modification of the Pd thin films can give rise to influential catalytic active positions, which are crucial for increasing oxygen reduction efficiency. Additionally, the bimetallic nanostructure of Au-modified Pd samples resulted in brilliant ORR catalytic ability. The synergistic interplays between Pd and Au positions may contribute to the better ORR performance of Au-modified Pd samples. Besides, improved ORR durability was delivered by the prepared 0.16 nm Au/Pd thin films. The boosted electro-catalytic activity and durability in the binary Au-modified Pd thin films offer an alternative option for Pt-free, highly durable oxygen reduction materials. The same structures and circumstances can activate Au to be utilized for various electrochemical reactions.

**CRediT authorship contribution statement**

**Fereshteh Dehghani Sanij**: Conceptualization, Methodology, Investigation, Formal analysis, Writing - original draft.

**Vitalii Latyshev**: Investigation, Methodology.

**Serhii Vorobiov**: Investigation.

**Hoydoo You**: Investigation, Formal analysis.

**Dominik Volavka**: Investigation, Formal analysis.

**Tomas Samuely**: Investigation, Formal analysis.

**Vladimir Komanicky**: Conceptualization, Funding acquisition, Supervision, Writing - review & editing.

**Conflict of interest**

The authors declare no conflict of interest in this work.

**Acknowledgement**

We thank Professor Kawaguchi, Tohoku U, for the useful discussion on X-ray reflectivity analysis. This work was supported by grants from the Slovak Research and Development Agency under contract APVV-20-0528 and the National Scholarship Programme of the Slovak Republic (NSP). The X-ray measurements performed at Argonne (HY) were supported by the U.S. Department of Energy (DOE), Office of Basic Energy Science (BES), Materials Sciences and Engineering Division under contract no. DE-AC02-06CH11357.

**References**

[1] Shao M, Chang Q, Dodelet JP, Chenitz R. Recent Advances in Electrocatalysts for Oxygen Reduction Reaction. Chem Rev 2016;116:3594–657. https://doi.org/10.1021/acs.chemrev.5b00462.

[2] Bhuvanendran N, Ravichandran S, Xu Q, Maiyalagan T, Su H. A quick guide to the assessment of key electrochemical performance indicators for the oxygen reduction reaction: A comprehensive review. Int J Hydrogen Energy 2022;47:7113–38. https://doi.org/10.1016/j.ijhydene.2021.12.072.

[3] Banham D, Ye S. Current status and future development of catalyst materials and catalyst layers for proton exchange membrane fuel cells: An industrial perspective. ACS Energy Lett 2017;2:629–38. https://doi.org/10.1021/acsenergylett.6b00644.

[4] Zaman S, Huang L, Douka AI, Yang H, You B, Xia BY. Oxygen Reduction Electrocatalysts toward Practical Fuel Cells: Progress and Perspectives. Angew Chemie - Int Ed 2021;60:17832–52. https://doi.org/10.1002/anie.202016977.

[5] Sanij FD, Balakrishnan P, Leung P, Shah A, Su H, Xu Q. Advanced Pd-based nanomaterials for electro-catalytic oxygen reduction in fuel cells: A review. Int J Hydrogen Energy 2021;46:14596–627. https://doi.org/10.1016/j.ijhydene.2021.01.185.

[6] Yang Y, Chen G, Zeng R, Villarino AM, Disalvo FJ, Van Dover RB, et al. Combinatorial Studies of Palladium-Based Oxygen Reduction Electrocatalysts for Alkaline Fuel Cells. J Am Chem Soc 2020;142:3980–8. https://doi.org/10.1021/jacs.9b13400.

[7] Zhou M, Guo J, Fang J. Nanoscale Design of Pd-Based Electrocatalysts for Oxygen Reduction Reaction Enhancement in Alkaline Media. Small Struct 2022;3:2100188. https://doi.org/10.1002/sstr.202100188.

[8] Lüsi M, Erikson H, Piirsoo HM, Paiste P, Aruväli J, Kikas A, et al. Oxygen reduction reaction on PdM/C (M = Pb, Sn, Bi) alloy nanocatalysts. J Electroanal Chem 2022;917. https://doi.org/10.1016/j.jelechem.2022.116391.

[9] Liu M, Xiao X, Li Q, Luo L, Ding M, Zhang B, et al. Recent progress of electrocatalysts for oxygen reduction in fuel cells. J Colloid Interface Sci 2022;607:791–815. https://doi.org/10.1016/j.jcis.2021.09.008.

[10] Dehghani Sanij F, Balakrishnan P, Su H, Khotseng L, Xu Q. Fabrication of polyoxometalate-modified palladium-nickel/reduced graphene oxide alloy catalysts for enhanced oxygen reduction reaction activity. RSC Adv 2021;11:39118–29. https://doi.org/10.1039/d1ra06936e.

[11] Rivera-Gavidia LM, Luis-Sunga M, Rodríguez JL, Pastor E, García G. Methanol tolerant Pd-Based carbon supported catalysts as cathode materials for direct methanol fuel cells. Int J Hydrogen Energy 2020;45:20673–8. https://doi.org/10.1016/j.ijhydene.2020.01.167.

[12] Dehghani Sanij F, Gharibi H. Preparation of bimetallic alloyed palladium-nickel electro-catalysts supported on carbon with superior catalytic performance towards oxygen reduction reaction. Colloids Surfaces A Physicochem Eng Asp 2018;538:429–42. https://doi.org/10.1016/j.colsurfa.2017.11.009.

[13] Sui S, Wang X, Zhou X, Su Y, Riffat S, Liu C jun. A comprehensive review of Pt electrocatalysts for the oxygen reduction reaction: Nanostructure, activity, mechanism

and carbon support in PEM fuel cells. J Mater Chem A 2017;5:1808–25. https://doi.org/10.1039/C6TA08580F.

[14] Xiao D, Jiang Q, Xu C, Yang C, Yang L, He H, et al. Interfacial engineering of worm-shaped palladium nanocrystals anchored on polyelectrolyte-modified MXene nanosheets for highly efficient methanol oxidation. J Colloid Interface Sci 2022;616:781–90. https://doi.org/10.1016/j.jcis.2022.02.111.

[15] Stamenkovic V, Mun BS, Mayrhofer KJJ, Ross PN, Markovic NM, Rossmeisl J, et al. Changing the activity of electrocatalysts for oxygen reduction by tuning the surface electronic structure. Angew Chemie - Int Ed 2006;45:2897–901. https://doi.org/10.1002/anie.200504386.

[16] Xiao Z, Wu H, Zhong H, Abdelhafiz A, Zeng J. De-alloyed PtCu/C catalysts with enhanced electrocatalytic performance for the oxygen reduction reaction. Nanoscale 2021;13:13896–904. https://doi.org/10.1039/d1nr02820k.

[17] Jiao W, Chen C, You W, Zhao X, Zhang J, Feng Y, et al. Hollow Palladium-Gold Nanochains with Periodic Concave Structures as Superior ORR Electrocatalysts and Highly Efficient SERS Substrates. Adv Energy Mater 2020;10:1–13. https://doi.org/10.1002/aenm.201904072.

[18] Wei Y-C, Wang K-W. Structure Alteration and Activity Improvement of PdCoAu/C Catalysts for the Oxygen Reduction Reaction. ECS Meet Abstr 2011;MA2011-02:885–885. https://doi.org/10.1149/ma2011-02/16/885.

[19] Zhang J, Sasaki K, Sutter E, Adzic RR. Stabilization of platinum oxygen-reduction electrocatalysts using gold clusters. Science (80- ) 2007;315:220–2. https://doi.org/10.1126/science.1134569.

[20] Takahashi S, Chiba H, Kato T, Endo S, Hayashi T, Todoroki N, et al. Oxygen reduction reaction activity and structural stability of Pt-Au nanoparticles prepared by arc-plasma deposition. Phys Chem Chem Phys 2015;17:18638–44. https://doi.org/10.1039/c5cp02048d.

[21] Zhang Y, Huang Q, Zou Z, Yang J, Vogel W, Yang H. Enhanced durability of Au cluster decorated Pt nanoparticles for the oxygen reduction reaction. J Phys Chem C 2010;114:6860–8. https://doi.org/10.1021/jp100559g.

[22] Kodama K, Jinnouchi R, Takahashi N, Murata H, Morimoto Y. Activities and Stabilities of Au-Modified Stepped-Pt Single-Crystal Electrodes as Model Cathode Catalysts in Polymer Electrolyte Fuel Cells. J Am Chem Soc 2016;138:4194–200. https://doi.org/10.1021/jacs.6b00359.

[23] Xu JB, Zhao TS, Shen SY, Li YS. Stabilization of the palladium electrocatalyst with alloyed gold for ethanol oxidation. Int J Hydrogen Energy 2010;35:6490–500. https://doi.org/10.1016/j.ijhydene.2010.04.016.

[24] Jiao W, Chen C, You W, Chen G, Xue S, Zhang J, et al. Tuning strain effect and surface composition in PdAu hollow nanospheres as highly efficient ORR electrocatalysts and SERS substrates. Appl Catal B Environ 2020;262:118298. https://doi.org/10.1016/j.apcatb.2019.118298.

[25] Xu JB, Zhao TS, Liang ZX. Synthesis of active platinum-silver alloy electrocatalyst toward the formic acid oxidation reaction. J Phys Chem C 2008;112:17362–7. https://doi.org/10.1021/jp8063933.

[26] Zhu F, Ma G, Bai Z, Hang R, Tang B, Zhang Z, et al. High activity of carbon nanotubes

[26] supported binary and ternary Pd-based catalysts for methanol, ethanol and formic acid electro-oxidation. J Power Sources 2013;242:610–20. https://doi.org/10.1016/j.jpowsour.2013.05.145.

[27] Shen X, Dai S, Pan Y, Yao L, Yang J, Pan X, et al. Tuning Electronic Structure and Lattice Diffusion Barrier of Ternary Pt-In-Ni for Both Improved Activity and Stability Properties in Oxygen Reduction Electrocatalysis. ACS Catal 2019;9:11431–7. https://doi.org/10.1021/acscatal.9b03430.

[28] R. Adzic. Electrocatalysis. Wiley-VCH, Weinheim, Germany; 1998.

[29] Wang J, Li M, Zhang J, Yan Y, Qiu X, Cai B, et al. Atom-Ratio-Conducted Tailoring of PdAu Bimetallic Nanocrystals with Distinctive Shapes and Dimensions for Boosting the ORR Performance. Chem - A Eur J 2020;26:4480–8. https://doi.org/10.1002/chem.201905284.

[30] He L-L, Song P, Feng J-J, Huang W-H, Wang Q-L, Wang A-J. Simple wet-chemical synthesis of alloyed PdAu nanochain networks with improved electrocatalytic properties. Electrochim Acta 2015;176:86–95. https://doi.org/10.1016/j.electacta.2015.06.137.

[31] Trindell JA, Duan Z, Henkelman G, Crooks RM. $Au_xPd(300-x)$ Alloy Nanoparticles for the Oxygen Reduction Reaction in Alkaline Media. ChemElectroChem 2020;7:3824–31. https://doi.org/10.1002/celc.202000971.

[32] Huang L, Han Y, Dong S. Highly-branched mesoporous Au-Pd-Pt trimetallic nanoflowers blooming on reduced graphene oxide as an oxygen reduction electrocatalyst. Chem Commun 2016;52:8659–62. https://doi.org/10.1039/c6cc03073d.

[33] Lv JJ, Li SS, Wang AJ, Mei LP, Chen JR, Feng JJ. Monodisperse Au-Pd bimetallic alloyed nanoparticles supported on reduced graphene oxide with enhanced electrocatalytic activity towards oxygen reduction reaction. Electrochim Acta 2014;136:521–8. https://doi.org/10.1016/j.electacta.2014.05.138.

[34] Lin XX, Zhang XF, Wang AJ, Fang KM, Yuan J, Feng JJ. Simple one-pot aqueous synthesis of AuPd alloy nanocrystals/reduced graphene oxide as highly efficient and stable electrocatalyst for oxygen reduction and hydrogen evolution reactions. J Colloid Interface Sci 2017;499:128–37. https://doi.org/10.1016/j.jcis.2017.03.087.

[35] Sarapuu A, Kasikov A, Laaksonen T, Kontturi K, Tammeveski K. Electrochemical reduction of oxygen on thin-film Pt electrodes in acid solutions. Electrochim Acta 2008;53:5873–80. https://doi.org/10.1016/j.electacta.2008.04.003.

[36] Sarapuu A, Hussain S, Kasikov A, Pollet BG, Tammeveski K. Electroreduction of oxygen on Nafion®-coated thin platinum films in acid media. J Electroanal Chem 2019;848:113292. https://doi.org/10.1016/j.jelechem.2019.113292.

[37] Wang C, Chi M, Li D, Strmcnik D, Van Der Vliet D, Wang G, et al. Design and synthesis of bimetallic electrocatalyst with multilayered Pt-skin surfaces. J Am Chem Soc 2011;133:14396–403. https://doi.org/10.1021/ja2047655.

[38] Erikson H, Antoniassi RM, Solla-Gullón J, Torresi RM, Tammeveski K, Feliu JM. Oxygen electroreduction on small (<10 nm) and {100}-oriented Pt nanoparticles. Electrochim Acta 2022;403:139631. https://doi.org/10.1016/j.electacta.2021.139631.

[39] Luo BJ, Wang L, Mott D, Njoki PN, Lin Y, He T, et al. Core / Shell Nanoparticles as Electrocatalysts for Fuel Cell Reactions ** 2008:4342–7. https://doi.org/10.1002/adma.200703009.

[40] Sarapuu A, Kasikov A, Wong N, Lucas CA, Sedghi G, Nichols RJ, et al. Electroreduction of oxygen on gold-supported nanostructured palladium films in acid solutions. Electrochim Acta 2010;55:6768–74. https://doi.org/10.1016/j.electacta.2010.05.092.

[41] Erikson H, Liik M, Sarapuu A, Marandi M, Sammelselg V, Tammeveski K. Electrocatalysis of oxygen reduction on electrodeposited Pd coatings on gold. J Electroanal Chem 2013;691:35–41. https://doi.org/10.1016/j.jelechem.2012.12.018.

[42] Chen D, Li J, Cui P, Liu H, Yang J. Gold-catalyzed formation of core-shell gold-palladium nanoparticles with palladium shells up to three atomic layers. J Mater Chem A 2016;4:3813–21. https://doi.org/10.1039/c5ta10303g.

[43] Hu Q, Zhan W, Guo Y, Luo L, Zhang R, Chen D, et al. Heat treatment bimetallic PdAu nanocatalyst for oxygen reduction reaction. J Energy Chem 2020;40:217–23. https://doi.org/10.1016/j.jechem.2019.05.011.

[44] Chen D, Li C, Liu H, Ye F, Yang J. Core-shell Au@Pd nanoparticles with enhanced catalytic activity for oxygen reduction reaction via core-shell Au@Ag/Pd constructions. Sci Rep 2015;5:1–9. https://doi.org/10.1038/srep11949.

[45] Xu Q, Chen W, Yan Y, Wu Z, Jiang Y, Li J, et al. Multimetallic AuPd@Pd@Pt core-interlayer-shell icosahedral electrocatalysts for highly efficient oxygen reduction reaction. Sci Bull 2018;63:494–501. https://doi.org/10.1016/j.scib.2018.03.013.